\documentclass[twocolumn,superscriptaddress,showpacs,amsmath,amssymb]{revtex4}
\usepackage{epsfig}
\usepackage{amsmath}
\usepackage{amssymb}
\usepackage{bm}

\DeclareMathAlphabet{\bi}{OML}{cmm}{b}{it}

\begin{document}
\title{Energy-Momentum dispersion relation of plasmarons in bilayer graphene}
\author{P. M. Krstaji\'c}
\affiliation{Institute of Microelectronic Technologies and Single Crystals (IHTM), University of Belgrade,
Njego\v{s}eva 12, 11000 Belgrade, Serbia}
\author{F. M. Peeters}
\affiliation{Departement Fysica, Universiteit Antwerpen, Groenenborgerlaan 171, B-2020 Antwerpen, Belgium}

\begin{abstract}
The relation between the energy and momentum of plasmarons in bilayer graphene is investigated within the Overhauser approach,
where the electron-plasmon interaction is described as a field theoretical
problem. We find that the Dirac-like spectrum is shifted by $\Delta E(\mathbf{k})\sim 100\div150\,{\rm meV}$
depending on the electron concentration $n_{e}$ and electron momentum. The shift increases with electron concentration
as the energy of plasmons becomes larger. The dispersion of plasmarons is more pronounced than in the case of single layer graphene, which is explained by the fact that the energy dispersion of electrons is quadratic and not linear.
We expect that these predictions can be verified using angle-resolved photoemission spectroscopy (ARPES).
\end{abstract}
\pacs{73.22.Pr, 73.20.Mf, 71.10.-w}
\maketitle

\vspace{5mm}
\section{Introduction}\label{intro} 
Coulomb interaction and plasmarons in both single layer graphene\cite{plasmaron1,plasmaron2,plasmaron3,plasmaron_science}
and bilayer graphene\cite{Sabas,Sensarma} have attracted recently a lot of interest. One of the reasons is that it was found experimentally that, for instance in monolayer graphene\cite{plasmaron_science} the accepted view of linear
(Dirac-like) spectrum does not provide a sufficiently accurate picture of the charge carrying excitations in this material.
The concept of a quasiparticle named "plasmaron", was introduced which is in fact a bound state of charge carriers with plasmons. The motivation behind the interest in this kind of studies is that exploring the physics of interaction
between electrons and plasmons may lead to realizations of "plasmonic" devices which merge photonics and electronics.
In earlier experiments, this more complicated picture of the band structure was not observed
because of the low quality and low mobility of old samples. The interest in similar phenomena in bilayer graphene is equally high.

Coulomb interaction and electronic screening was probed in bilayer and multilayer graphene using angle-resolved photoemission
spectroscopy (ARPES) in Ref.~\onlinecite{BostwickPRL}. Recently, Sensarma \textit{et al.}\cite{Sensarma} investigated plasmarons and the quantum spectral function in bilayer graphene theoretically. The authors of that reference predicted a broad plasmaron peak away from
the Fermi surface. Similar findings were reported in Ref.~\onlinecite{Sabas} where thermal Green's functions in both random phase
approximation and self-consistent GW approximation were used to determine the spectral function. Bilayer
graphene shares some properties with both graphene and the two-dimensional electron gas found in common semiconductors. While
the energy dispersion in graphene is linear in momentum, in bilayer graphene it is nearly quadratic. The advantage of bilayer
graphene over usual semiconductors is that its charge carrier density can be controlled by the application of a gate voltage
over orders of magnitude and this from electrons to holes.
Furthermore, the band gap can be tuned to meet requirements for several device applications.

In this paper, we employ an alternative approach in order to determine the energy spectrum of bilayer graphene. The approach is based on second order perturbation theory of the electron-plasmon interaction and the problem is cast into a field theoretical problem. By doing so, one is able to evaluate the correction to the band structure which comes as a consequence of the interaction of charge carriers with plasmons.
As far as the interaction between plasmons and charge carriers is concerned, we generalize the Overhauser approach\cite{Overhauser,remotepol} to the two-dimensional electron gas in bilayer graphene.

We organize the paper as follows. In Sec.~\ref{theory} we present the theoretical model and give pertinent expressions for the interaction and the coupling between electrons and plasmons in bilayer graphene. In the subsequent section, Sec.~\ref{results}, the numerical calculations of the energy correction due to the interaction with plasmons are presented for
various doping levels, i.e. charge carrier density. The influence of the doping level is analyzed and discussed. Finally, we summarize our results and present the conclusions in Sec.~\ref{summary}.

\section{Theoretical model}\label{theory}
If the relevant energy scale in bilayer graphene is smaller than the interlayer hopping parameter $t_{\perp}$, one may use the low energy limit.
In this limit, the problem can be reduced to the effective two-band model and the corresponding Hamiltonian reads\cite{Kotov}
\begin{equation}\label{Ham0}
\mathcal{H}_{0}=-\frac{v_{F}^{2}}{t_{\perp}}\left(\begin{array}{cc}
0 & \pi^{\dagger2} \\
\pi^{2} & 0 \end{array}\right)\,,
\end{equation}
\noindent where $v_{F}$ is the Fermi velocity and $\pi=p_{x}+ip_{y}$. The eigenvalues of Eq.~(\ref{Ham0}) are well known and read $E^{(0)}_{s}=s\hbar^{2}k^{2}/2m_{e}$ where $s=\pm1$.
Here we introduced the effective mass $m_{e}=t_{\perp}/(2v_{F}^{2})$ in the low energy limit, and $m_{e}\approx0.034m_{0}$.
Please note that we are interested in energies larger than $1-5 {\rm meV}$ such that the usual warping is not important. It is well known that graphene structures may sustain quanta of collective charge excitations of the electron gas,
i.e. plasmons, due to the restoring force of the long-range $1/r$ Coulomb interaction. However in contrast to the usual
two-dimensional electron gas, the "Dirac plasma" is manifestly of quantum nature\cite{SarmaPRL}. For instance, in single layer
 graphene the plasma frequency is proportional to $1/\sqrt{\hbar}$, and does not have a classical limit independent of the
 Planck constant. As far as bilayer graphene is concerned the plasma frequency (in the long wavelength limit) is given by\cite{SarmaRev}
\begin{equation}
\omega_{\mathbf{q}}=\left(\frac{2\pi n_{e}e^2}{\kappa m_{e}}\right)^{\frac{1}{2}}\sqrt{q}\,,
\end{equation}
\noindent where $\kappa$ is the dielectric constant of the material and is related to the one of the substrate, $\kappa=(1+\kappa_{s})/2\approx2.5$ for ${\rm SiO}_{2}$ substrate. The excitations of the electron gas are represented by a scalar field previously described by Overhauser\cite{Overhauser} for the $3$D electron gas.
The correction in the electron spectrum are calculated analogously as for the polaron problem where now a test charge interacts with the plasmon field. The interaction of an electron displaced from
 the graphene layer with plasmons was treated in our earlier work\cite{remotepol}, and the interaction term of the Hamiltonian is given by
\begin{equation}
H_{int}=\sum_{\mathbf{q}}\frac{V_{\mathbf{q}}}{\sqrt{\Omega}}\exp(i\mathbf{q}\cdot\mathbf{r})(a_{\mathbf{q}}+a_{-\mathbf{q}}^{\dagger})\,,
\end{equation}
\noindent where the electron-plasmon interaction matrix element is\cite{PeetPol}
\begin{equation}\label{Vq1}
V_{\mathbf{q}}=\frac{2\pi e^{2}}{\sqrt{\Omega}\kappa q}\lambda_{\mathbf{q}}\,.
\end{equation}
 Its value can be determined using the $f$-sum rule applied to the case of interest. The starting point is the fact that the expectation value of the double commutator $\langle n|[n_{-\mathbf{q}},[n_{\mathbf{q}},H]]|0\rangle$ can be evaluated in two different ways\cite{Pines}. First, it is known that the relation
$\langle n|C|m\rangle=(E_{n}-E_{m})\langle n|A|m\rangle$ holds for an arbitrary commutator with the Hamiltonian, $C=[H,A]$. Second,
it can easily be proven that
\begin{equation}\label{Vq2}
\langle n|[n_{-\mathbf{q}},[n_{\mathbf{q}},H]]|0\rangle=2\sum_{n}\hbar\omega_{n0}|\langle n|n_{\mathbf{q}}|0\rangle|^{2}\,,
\end{equation}
\noindent where $\hbar\omega_{n0}=E_{n}-E_{0}$. Further, the explicit evaluation of the double commutator yields
\begin{equation}\label{Vq3}
\sum_{n}\hbar\omega_{n0}|\langle n|n_{\mathbf{q}}|0\rangle|^{2}=N\frac{\hbar^{2}q^{2}}{2m_{e}}\,.
\end{equation}
The $f$-sum rule then reduces to
\begin{equation}\label{Vq4}
\hbar\omega_{\mathbf{q}}^{\prime}\lambda_{\mathbf{q}}^{2}=N\frac{\hbar^{2}q^{2}}{2m_{e}}\,.
\end{equation}
On combining relations in Eqs.~(\ref{Vq1}-\ref{Vq4}) we arrive at the expression for the interaction matrix element
\begin{equation}\label{eqVq}
V_{\mathbf{q}}=\frac{2\pi e^{2}}{\kappa}\sqrt{\frac{\hbar n_{e}}{2m_{e}\omega_{\mathbf{q}}^{\prime}}}\,,
\end{equation}
\noindent where $n_{e}$ is the electron concentration, $n_{e}=N/\Omega$. Note that $\omega_{\mathbf{q}}^{\prime}$ is not the bare plasmon frequency
but is altered by the polarization of the electron gas. We need the dielectric function of the electron in order
to determine $\omega_{\mathbf{q}}^{\prime}$. It can be shown that
in the long wavelength limit ($q\rightarrow0$) and within the random phase approximation (RPA) the dielectric function can be approximated by the following relation\cite{dielSarma}
\begin{equation}\label{epsqE2}
\epsilon(q)=1+\frac{q_s}{q}\,,
\end{equation}
\noindent where $q_{s}$ is the screening wavevector\cite{SarmaRev} and given by $q_{s}=2\pi e^{2}/\kappa D_{0}$ while $D_{0}$ is the density of states of bilayer graphene, $D_{0}=g_{s}g_{v}m_{e}/(2\pi\hbar^{2})$. Here $g_{s}$ and $g_{v}$ are the degeneracy factors for spin and valley degrees of freedom. The previous relation, Eq.~(\ref{epsqE2}) is obtained from the general relation
$\epsilon(q,\omega)=1+v_{c}(q)\Pi(q,\omega)$, where $v_{c}(q)=2\pi e^2/(\kappa q)$ is the Fourier transform of the two-dimensional Coulomb interaction, and $\Pi(q,\omega)$ is the $2$D polarizability. Finally, the actual plasmon frequency is given by\cite{remotepol}
\begin{equation}\label{epsqE1}
\omega_{\mathbf{q}}^{\prime2}=\omega_{\mathbf{q}}^2\frac{\epsilon(q)}{\epsilon(q)-1}\,.
\end{equation}

Now, we are ready to evaluate the correction in the energy spectrum due to the interaction between electrons and plasmons. This will be carried out by employing second order perturbation theory,
and for the case of bilayer graphene it reads

\begin{equation}\label{corrE}
\Delta E_{0}(\mathbf{k})=-P\frac{1}{\Omega}\sum_{\mathbf{q}}\frac{|V_{\mathbf{q}}|^2}{\hbar\omega_{\mathbf{q}}+E_{0}(\mathbf{k}-\mathbf{q})-E_{0}(\mathbf{k})}\,,
\end{equation}
where $P(\cdot)$ stands for the principal value. The cut-off value for the momentum $\mathbf{q}$ was taken to be $q_{c}=1/a_{0}$ where $a_{0}$ is the lattice constant. Note that this correction is given within non-degenerate Rayleigh-Schr\"{o}dinger perturbation theory (RSPT). However, for certain values of the plasmon wavevector $\mathbf{q}$ a degeneracy occurs when $E_{0}(\mathbf{k})=\hbar\omega_{\mathbf{q}}+E_{0}(\mathbf{k}-\mathbf{q})$. Because of this degeneracy, improved Wigner
Brillouin perturbation theory\cite{PeetPol} (IWBPT) can be employed to tackle this problem. The main idea behind this method is to ensure enhanced convergence when the denominator in Eq.~(\ref{corrE}) approaches zero, which is realized by adding the term $\Delta(\mathbf{k}) = \Delta E(\mathbf{k})-\Delta E_{0}(\mathbf{k})$ ($\Delta E_{0}(\mathbf{k})$ is evaluated within RSPT),

\begin{equation}\label{corrEIW}
\Delta E(\mathbf{k})=-P\sum_{\mathbf{q}}\frac{|V_{\mathbf{q}}|^2}{\hbar\omega_{\mathbf{q}}+E_{0}(\mathbf{k-q})-
E_{0}(\mathbf{k})-\Delta(\mathbf{k})}.
\end{equation}

Equation.~(\ref{corrEIW}) should be solved self-consistently since $\Delta E$ appears on both sides of the equation. Note that
$E(\mathbf{k})=E(k)$ due to the isotropic nature of the spectrum. In the following section the value of $\Delta E(k)$
will be calculated numerically for concrete values of the doping level, permittivity and other parameters of the
material. As pointed out in Ref.~\onlinecite{plasmaron3} the plasmon excitation in graphene of the Dirac sea remains pretty much
well defined even when it penetrates the interband particle-hole continuum. This is because the transitions near the bottom of
the interband particle-hole continuum have almost parallel wavevectors $\mathbf{k}$ and $\mathbf{k+q}$ and therefore carry negligible charge-fluctuation weight.
A similar conclusion holds for bilayer graphene. In practice, the damping can be important for very large momentum $q$,
but then the contribution to the energy shift, i.e. to the integral in Eq.~(\ref{corrEIW}) is small.

\begin{figure}[t]
\includegraphics[width=8cm]{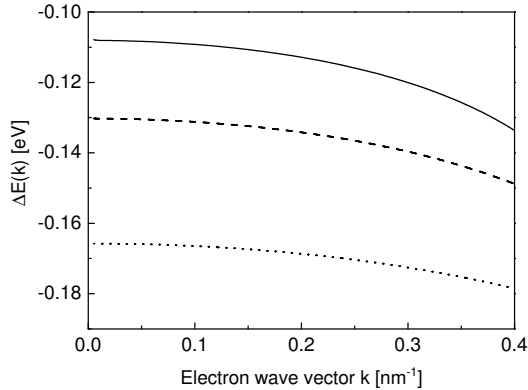}
\caption{\label{fig1}The correction to the energy, $\Delta E(k)$, vs electron momentum $k$,
for three values of the doping level $n_{e}=3\times10^{12}\,{\rm cm}^{-2}$ (solid curve), $5\times10^{12}\,{\rm cm}^{-2}$
(dashed curve) and $10^{13}\,{\rm cm}^{-2}$ (dotted curve).}
\end{figure}

\section{Numerical results}\label{results}
The numerical calculations are carried out for doped bilayer graphene, with varying electron concentration.
First we give in Fig.~\ref{fig1} the results for the energy correction
$\Delta E(\mathbf{k})$ for three levels of doping\cite{comment44}: $n_{e}=3\times10^{12} \,{\rm cm}^{-2}$ (solid curve),
$5\times10^{12}\,{\rm cm}^{-2}$ (dashed curve) and $10^{13} \,{\rm cm}^{-2}$ (dotted curve).
As can be seen from the figure, the shift increases with the electron momentum.
The increase with $\mathbf{k}$ is more rapid than in the case of single layer graphene\cite{Krstajplasm}. 
This is qualitatively similar to what was found in Refs.~\onlinecite{Sabas} and \onlinecite{Sensarma} where a broad peak was attributed to plasmarons. While the explicit
dependence on the concentration is the same $V_{\mathbf{q}}\propto\sqrt{n_{e}}$, the interaction matrix element is related to
the doping level also through the plasmon frequency. The latter in single layer graphene is mainly proportional
to $n_{e}^{1/4}$ while in bilayer graphene is $\sqrt{n_{e}}$. Further, the effective plasmon frequency is modulated
through the polarization of the surrounding electron gas. One should not forget that a property of bilayer graphene, important for the present analysis, is the fact that the coupling parameter is a function of
the carrier density $r_{s}\propto n_{e}^{-1/2}$ (while in single layer graphene it is independent of $n_{e}$).
More precisely, the strength of the Coulomb interaction is tunable and depends on the level of doping. As for the comparison with earlier theoretical findings, we got $\Delta E(k=k_{F})=0.12 {\rm eV}$ for $n_{e}=3\times10^{12} \,{\rm cm}^{-2}$
while the authors of Ref.~\onlinecite{Sensarma} obtained the value $\Delta E(k=k_{F})=0.18 {\rm eV}$.
\begin{figure}[t]
\includegraphics[width=8cm]{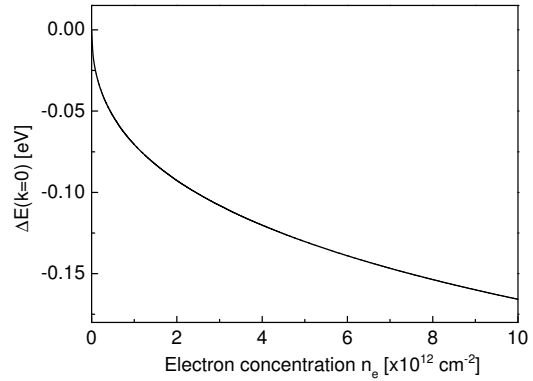}
\caption{\label{fig2}The correction to the energy $\Delta E(0)$ for $k=0$, vs the electron concentration $n_{e}$.}
\end{figure}

Unlike the case of polarons in conventional semiconductors, here it is not straightforward to derive
any approximate analytical relation for $\Delta E(k)$ at small $k$. This is due to the fact that plasmons here
have a more complicated dispersion relation, and the fact that the interaction strength $V_{\mathbf{q}}$
depends on $\mathbf{q}$ in a non-trivial manner.
Thus we will treat Eq.~(\ref{corrE}) numerically and one may write for small $k$
\begin{equation}\label{apprdE}
\Delta E(k)=\Delta E(0)+\alpha k^2+\beta k^4.
\end{equation}
We fitted Eq.~(\ref{apprdE}) to our numerical results within the range $0<k<0.5 {\rm nm}^{-1}$. For instance, for $n_{e}=3\times10^{12}\,{\rm cm}^{-2}$ the fitting parameters are $\alpha=-9.09\times10^{-16}\,{\rm eVcm}$
and $\beta=-4.73\times10^{-29}\,{\rm eVcm}^{2}$.
\begin{figure}[h]
\includegraphics[width=8cm]{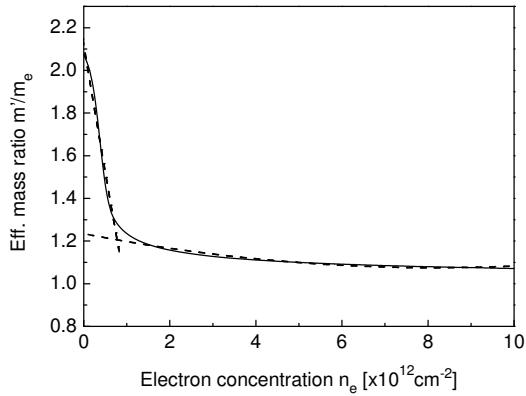}
\caption{\label{fig3} Ratio between the effective mass of the plasmaron band and the electron as a function of
the electron concentration $n_{e}$. The dashed curves show the low and high density asymptotics.}
\end{figure}

In Fig.~\ref{fig2} we present the result for the energy correction $\Delta E(0)$ at $k=0$, vs doping of bilayer graphene, i.e. the electron
concentration $n_{e}$. As can be seen, the absolute value of $\Delta E(0)$ increases with the electron concentration. This is mainly the
consequence of the dependence of the matrix element $V_{\mathbf{q}}$ on $n_{e}$ (see Eqs.~(\ref{eqVq}) and (\ref{epsqE1})).
As mentioned earlier the relation is complicated since the plasmon frequency is modified through the polarization of the electron gas.
We note that the obtained results for the energy shift on the concentration can be fitted (for $0<n_{e}<10^{13} {\rm cm}^{-2}$) to $\Delta E(0)=an_{e}^{\alpha}/(1+bn_{e}^{\gamma})$,
where $\alpha=0.53$, $\gamma=0.36$ and $a=-4.7\times10^{-8}$, $b=2.53\times10^{-5}$ ($n_{e}$ is given ${\rm cm}^{-2}$ and $\Delta E(0)$ in ${\rm eV}$). It would be instructive to determine the effective mass of the plasmaron band defined by $E(k)=E_{0}(k)+\Delta E(k)=\hbar^{2}k^{2}/2m^{\prime}$. Figure \ref{fig3} shows the dependence of the effective mass ratio $m^{\prime}/m_{e}$ on the electron concentration. The ratio starts
from a value around $2$ and and drops rapidly to $1.2$ while for $n_{e}>2\times 10^{12} {\rm cm}^{-2}$ it converges slowly to $1$. The low and high density behavior fits are shown by dashed lines in Fig.~\ref{fig3} and were fitted respectively to the following expressions: $2.1-1.2\times10^{-12}{\rm cm^{2}}\cdot n_{e}+5.1\times10^{-26}{\rm cm^{4}}\cdot n_{e}^{2}$ for $n_{e}\le8\times10^{11}{\rm cm}^{-2}$ and $1.2-3.9\times10^{-14}{\rm cm^{2}}\cdot n_{e}+2.4\times10^{-27}{\rm cm^{4}}\cdot n_{e}^{2}$, for  $n_{e}>8\times10^{11}{\rm cm}^{-2}$. Note that the large dispersion of the plasmaron band was also revealed in recent theoretical investigation of the spectral function\cite{Sensarma},
where it was determined that plasmarons do have a broad peak.

\section{Conclusion}\label{summary}
In this paper we investigated the interaction between an electron and the collective excitation of the electron gas,
i.e. plasmons, in bilayer graphene by using a field-theoretical approach. This interaction is modeled by generalizing
the Overhauser approach \cite{Overhauser} to the $2$DEG in this material. We evaluated the energy correction, that is
the shift in the energy spectrum as a result of this interaction. We employed second order perturbation theory in
order to determine the energy of the plasmaron, which is a composite quasi-particle, i.e. a bound state of an electron
with plasmons.

The motivation behind the present study are the increased interest in the spectral function of bilayer
graphene\cite{Sabas,Sensarma}, and the prediction of the existence of a broad plasmaron peak away\cite{Sensarma},
but near the Fermi surface. First we evaluated the correction to the energy as a result of the interaction
between electron and plasmons. The shift is appreciable and lies in the range
of $100\div150 {\rm meV}$ depending on the electron concentration and electron wavevector.

Further, we investigated the influence of the doping level on the shift $\Delta E(0)$,
and it is shown that it increases with $n_{e}$ which is more pronounced than in the case of single layer graphene\cite{Krstajplasm}.
The difference with single layer graphene lies in the actual dependence of the interaction strength $V_{\mathbf{q}}$
on the electron concentration. This should be revealed in future angle-resolved photoemission spectroscopy.
\acknowledgments{}
This work was supported by the Flemish Science Foundation (FWO-Vl), the ESF-EuroGRAPHENE project CONGRAN
and by the Serbian Ministry of Education and Science, within
the project No. TR~32008.


\begin{references}
\bibitem{comment44} When we replace the parabolic dispersion relation by the one obtained from the $4\times4$ Hamiltonian we found that the energy
correction increases with less than $8\%$.
\end{references}
\end{document}